# Anisotropy and magnetization reversal with chains of submicron-sized Co hollow spheres

Lin He and Chinping Chen[a]

[1]Department of Physics, Peking University, Beijing 100871, People's Republic of China

a) E-mail : cpchen@pku.edu.cn, Phone : +86-10-62751751

Fang Liang and Lin Guo[b]

[2]School of Materials Science and Engineering, Beijing University of Aeronautics and Astronautics, Beijing 100083, People's Republic of China

b) E-mail: guolin@buaa.edu.cn

Abstract

Magnetic properties with chains of hcp Co hollow spheres have been studied. The diameter of the spheres ranges from 500 to 800 nm, with a typical shell thickness of about 60 nm. The shell is polycrystalline with an average crystallite size of 20 to 35 nm. The blocking temperature determined by the zero-field-cooling $M_{\text{ZFC}}(T)$ measurement at $H$ = 90 Oe is about 325 K. The corresponding effective anisotropy is determined as, $K_{\text{eff}} \sim 4.6 \times 10^4$ J/m$^3$. In addition, the blocking temperature and the effective anisotropy determined by the analysis on $H_C(T)$ are 395 K and 5.7 ×10$^4$ J/m$^3$, respectively. The experimentally determined anisotropy is smaller by one order of magnitude than the magnetocrystalline anisotropy of the bulk hcp Co, which is about 3 to 5 ×10$^5$ J/m$^3$. A further analysis on $H_C(T)$ shows that the magnetization reversal follows a nucleation rotational mode with an effective switching volume, $V^* \sim 2.3$ ×10$^3$ nm$^3$. The corresponding effective diameter is calculated as 16.4 nm. It is slightly larger than the coherence length of Co, about 15 nm. The possible reason for the much reduced magnetic anisotropy is discussed briefly.

1. Introduction

As the synthesis technique in material science makes progress to reach the realms of nanometer, the emergence of nano-scaled magnetic material brings about a series of new applications due to its unique properties. Recently, much attention has been focused on the fabrication of nano- or submicron-sized hollow spheres because of their potential applications in catalysts, artificial cells, coatings, medical delivery vehicle systems for inks, and dyes [1-4], *etc*. The magnetic properties of the hollow sphere structure have been reported also. For example, the magnetization ground states of magnetic hollow sphere are studied by analytical consideration by D. Goll *et al*. [5]. Two stable ground states, the single domain and the vortex curling, have been obtained in theory for this special structure. Experimentally, there are many works concerning the synthesis and magnetic properties of Ni or Co hollow spheres [6-8]. However, these works mainly concentrate on the characterization of magnetic properties only. Therefore, an in-depth investigation is of great interest and importance.

In our previous works, we have reported the synthesis and characterization of uniform chains of Co hollow spheres [9] or Co hollow spheres with each cavity void enclosing a solid nanoparticle [10]. In the present paper, we report a study on the magnetic anisotropy properties for the chains of hcp Co hollow structure. The experimental details are presented in [9]. The blocking temperature obtained from the $M_{ZFC}(T)$ measurement is, $T_B^{ZFC}(90Oe) \sim 325$ K, whereas it is $T_B^{NB} \sim 395$ K determined by the Néel–Brown analysis on $H_C(T)$. The corresponding effective anisotropy constant is derived as $K_{eff} \sim 4.6 \times 10^4$ J/m$^3$ and $5.7 \times 10^4$ J/m$^3$, respectively. These values are one order smaller than the magnetocrystalline anisotropy, $K_1$, of the bulk hcp Co, which is about 3 to $5 \times 10^5$ J/m$^3$ [11-13].

2. Sample preparation and characterization

The preparation procedure and mechanism of synthesis are reported in [9]. In brief, the sample was synthesized by the reduction of CoCl$_2$·6H$_2$O with hydrazine monohydrate in the presence of poly(vinylpyrrolidone) (PVP, $M$w 58,000). First, a solution was prepared by dissolving cobalt salt and PVP in ethylene glycol (EG). This was followed by a dropwise addition of hydrazine monohydrate at room temperature with simultaneous vigorous agitation. Second, the homogeneous mixture was heated to the boiling point of EG for refluxing (197 $^o$C). After the mixture was refluxed for 4 h, a dark precipitate was obtained, which was separated by centrifugation and washed with absolute ethanol.

The sample has been characterized and presented in the previous report [9]. The major results are described as follows. The XRD spectrum is in agreement with the hcp Co (JCPDS 05-0727). No impurity phase such as the cobalt oxide or the precursor compounds have been detected. The structural analysis by high-resolution transmission electron microscopy (HRTEM) has also confirmed the hcp crystal structure [9]. Scanning electron microscopy (SEM) images of the sample are obtained by a JSM-5800, as shown in Fig. 1. The sample consists of cobalt hollow spheres with size ranging from 500 to 800 nm. The spheres are connecting with each other in structure, forming branched necklace-like chains with a length of tens of micrometers. The surface of these hollow spheres is rough. It reflects a polycrystalline structure formed of Co nanocrystallites. According to the detailed investigation by SEM and TEM, as presented in [9], there is a wide distribution in the size of crystallites. The average size is about 20-35 nm in diameter. The shell thickness is estimated to be about 60 nm. PVP plays a crucial role in the formation of chains of hollow sphere structure. It serves not only as the soft template, but also as the surface reagent. Without PVP, only Co solid spheres with irregular shape were obtained. At the end of the experiments, the residual PVP was washed away thoroughly.

3. Magnetic measurements and analysis

The temperature and field dependence of the magnetization, $M(T)$ and $M(H)$, was investigated by a Quantum Design SQUID magnetometer. In particular, $M(T)$ was studied by zero-field-cooling (ZFC) and field-cooling (FC) modes from 5 K to 380 K. For the $M_{ZFC}(T)$ curve, the sample was first cooled down to $T$ = 5 K under zero applied field, and then the magnetization signal was recorded in an applied field, $H_{app}$ = 90 Oe, in the warming process. For the $M_{FC}(T)$ curve, the procedure was the same, except that the sample was cooled under a cooling magnetic field, $H_{COOL}$ = 20 kOe.

In Fig. 2, the curves for $M_{ZFC}(T)$ and $M_{FC}(T)$ are presented. These two separate widely from each other. In addition, a maximum shows up on the $M_{ZFC}(T)$ curve, as shown in the inset of Fig. 2. The position of the maximum is defined as the blocking temperature, $T_B^{ZFC}$(90 Oe) ~ 325 K. These properties are attributed to the presence of a magnetic anisotropy barrier, $E_A$. Above $T_B^{ZFC}$, $M_{ZFC}(T)$ and $M_{FC}(T)$ still separate slightly from each other. This is attributable to the size distribution of switching volume, which in turn is dictated by the size of nanocrystallites forming the shell of sphere. The size of the switching volume in comparison with that of the nanocrystallites and the coherence length of Co will be elaborated later.

$M(H)$ curves at $T$ = 50, 70, 100, 150, 200, 250, 300, and 380 K were measured by the sweeping field within the limit from -50 kOe to 50 kOe. Fig. 3 shows three representative hysteresis loops in the low field region measured at $T$ = 50, 150 and 380 K. We see that the remanent magnetization and the coercivity decrease with the increasing temperature. In the high field region, however, the saturation magnetization $M_S(T)$ stays almost a constant at these temperatures. The inset shows the saturation behavior of $M(H)$ curve in the high field at $T$ = 300 K. The saturation value is determined as 168 emu/g, which is equal to the bulk value. This reveals a different feature from that of the recently reported nano-sized Ni hollow spheres with a much reduced saturation magnetization, ~ 24% [6] or ~ 38% [7] of the bulk value at 300 K.

Fig. 4 shows the temperature dependence of the reduced saturation magnetization, $M_S(T)/M_S(50K)$, and the remanent ratio, $M_R(T)/M_S(T)$. The saturation magnetization almost remains constant in the investigated temperature range, whereas the magnitude of $M_R/M_S$ decreases almost linearly with increasing temperature. The remanent ratio of the sample is about 0.12 at 50 K, and it goes down to about 0.02 at 380 K. This is much smaller than the theoretically analyzed value, *i.e.* 0.5, for a system of randomly oriented Stoner-Wohlfarth (SW) particles [14]. Hence, it is a sign that the magnetization reversal of the present sample is not by the coherent rotational mode.

4. Anisotropy and blocking temperature

The magnetic anisotropy and the magnetization reversal behavior are analyzed from the results of $M(T)$ and $M(H)$ measurements. In Fig. 5, the coercivity, $H_C(T)$, determined from the hysteresis loops is plotted along with $\Delta M(T) = M_{FC} - M_{ZFC}$. These two exhibit a proportional relation. The temperature dependent coercivity can be analyzed by the Néel–Brown model [15,16]. By this analysis, the field dependent anisotropy barrier is written as $E_A(H) = E_0(1-H/H_0)^\alpha$, where $E_0$ is the energy barrier at zero field and $H_0$ is the switching field at zero temperature. By taking into account the thermal activation effect over this barrier, it is then expressed as $E_A(H) = k_BT\ln[t(T)/t_0]$, in which $t$ is the time necessary to jump over the energy barrier at temperature $T$ and $t_0$ is a constant typically of the order from $10^{-9}$ to $10^{-11}$ s. Usually, $\ln[t(T)/t_0]$ equals to 25. The temperature dependent property of the coercivity then follows,

$$H_C(T) = H_0\left\{1 - \left[\frac{k_BT}{E_0}\ln(t/t_0)\right]^{1/\alpha}\right\}. \qquad (1)$$

For particles with size larger than the magnetic coherence length, by taking into account the intraparticle interaction including the exchange and the dipolar interactions between different magnetic domains and neglecting the interparticle interaction, the exponent $\alpha$ has been shown theoretically ~ 3/2 by R. H. Victora [17].

Critical length scales for the magnetic localization and the related magnetization reversal property have been discussed in detail by Skomski *et al*. for transition-metal nanowires, including single crystal and polycrystalline [18]. The crystallite size of the present polycrystalline sample is about 20 ~ 35 nm on average. It is larger than the coherence length of Co, ~ 15 nm [19]. This would put the sample in the ordinary weak coupling region in the phase diagram proposed by Skomski *et al.* for the polycrystalline nanowire, see Fig. 2 in [18], despite the much more complicated structure with the present sample. Therefore, $H_C(T)$ is expected to follow Eq. (1) with $\alpha = 3/2$ for the magnetization reversal by nucleation rotational mode. The dashed line in Fig. 5 is the fitting result of Eq. (1) by fixing $\alpha = 3/2$. It describes reasonably well the behavior of $H_C(T)$ and $\Delta M(T)$, as shown in Fig. 5. The values of the fitting parameters determined by Eq. (1) are $H_0 = 415$ Oe and $E_0/[k_B\ln(t/t_0)] = 395$ K. Thus, the blocking temperature obtained by the Néel–Brown analysis on $H_C(T)$ is $T_B^{NB} = 395$ K, which is larger than $T_B^{ZFC} = 325$ K. It reflects the blocking property corresponding to a switching volume size in the large end of the distribution. On the other hand, $T_B^{ZFC}$ describes the blocking property for the average switching volume size.

The effective switching volume is estimated as, $V^* \sim 2.3 \times 10^3$ nm$^3$, according to the expression $V^* = E_0/(\mu_0 M_S H_0)$ with $E_0$ and $H_0$ obtained from the fitting by Eq. (1) [20]. The corresponding effective diameter is about 16.4 nm. This is slightly smaller than the average diameter of the nanocrystallites, 20 ~ 35 nm [9]. Importantly, this value is consistent with the coherence length of Co, ~ 15 nm. This further supports the picture of magnetization reversal by nucleation rotation. With the knowledge of $T_B$ and $V^*$, the effective anisotropy, $K_{eff}$, can be calculated by $K_{eff}V^* = 25 k_B T_B$. It is about 4.6 ×10$^4$ J/m$^3$ derived from $T_B^{ZFC}$ or 5.7 ×10$^4$ J/m$^3$ from $T_B^{NB}$. This is one order smaller than the magnitude of magnetocrystalline anisotropy of the bulk hcp Co, $K_1 \sim$ 3 to 5 ×10$^5$ J/m$^3$ [11-13]. For large particles formed of nanocrystallites which are small with randomly oriented axes of anisotropy and in the strongly exchange coupled region, the average magnetocrystalline anisotropy becomes complicated [21]. In the present work, however, the shell thickness of about 60 nm is larger than the nanocrystallite size of 20 to 35 nm, which in turn is larger than the switching volume size of about 16.4 nm. The magnetization reversal for an individual grain proceeds by the nucleation rotational mode. As shown by Fig. 3 in [18], a localization of the nucleation mode would lead to a softening of magnetism in a single crystal nanowire. In the present work, we consider spherical grains rather than elongated ones with the nucleation rotational mode. Perhaps, this is the reason leading to the significantly reduced anisotropy.

5. Conclusion

In summary, the magnetic properties of chains of submicron-sized hcp Co hollow spheres have been investigated. The effective anisotropy is found to be one order smaller than the magnetocrystalline anisotropy of the bulk hcp Co. According to the Néel–Brown analysis on $H_C(T)$, the magnetization reversal is by nucleation rotational mode. The blocking temperature determined from the maximum of the $M_{ZFC}(T)$ curve, measured in the applied field $H_{app}$ = 90 Oe, is $T_B^{ZFC} \sim$ 325 K. This is lower than the one derived from the behavior of $H_C(T)$, $T_B^{NB} \sim$ 395 K. The difference between these two values is attributable to the size distribution of the switching volume, which in turn is affected by the size of the nanocrystallites forming the shell of hollow spheres.

Acknowledgement

Authors acknowledge the support from the National Natural Science Foundation of China (No. 20673009), the program for New Century Excellent Talents in University (NCET-04-0164) and the Specialized Research Fund for the Doctral Program of Higher Education (SRFDP-2006006005).

Figures

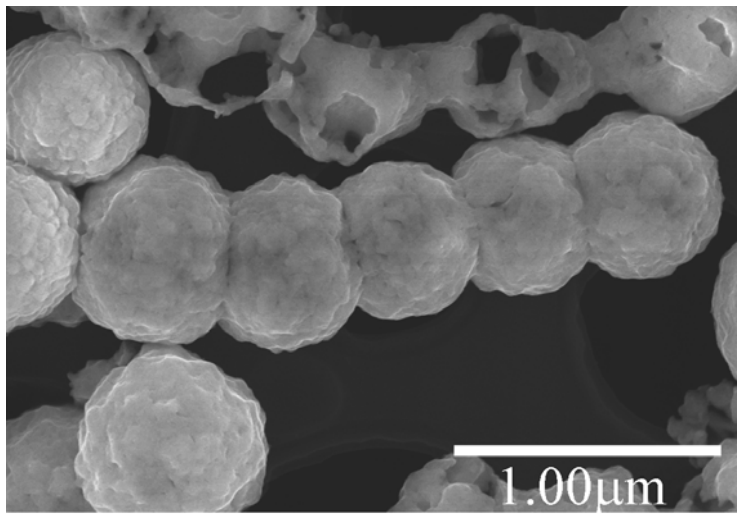

Figure 1: SEM image for the submicron-sized chains of Co hollow spheres.

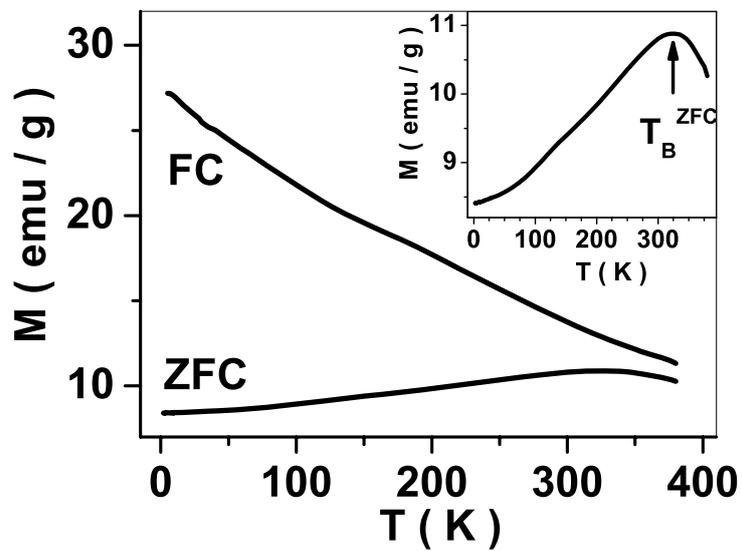

Figure 2: $M_{ZFC}(T)$ and $M_{FC}(T)$ curves measured in an applied field, $H_{app}$ = 90 Oe between 5 K and 380 K. The inset shows an enlarged view of the $M_{ZFC}(T)$ curve.

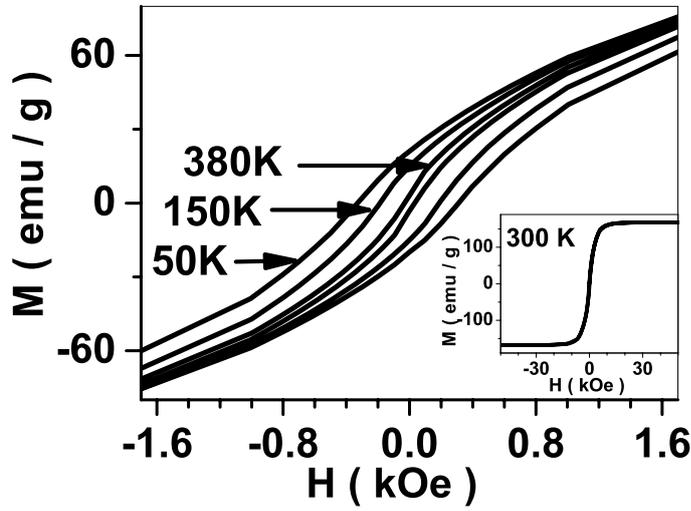

Figure 3: Three hysteresis loops measured at $T$ = 50 K, 150 K, and 380 K. The inset shows the saturation magnetization measured at 300 K.

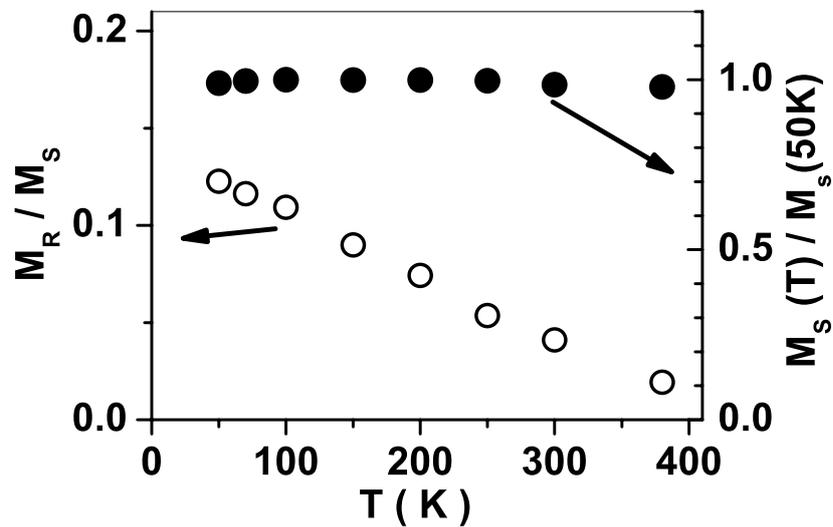

Figure 4: Reduced remanent ratio, $M_R(T)/M_S(T)$, and reduced saturation magnetization, $M_S(T)/M_S(50K)$, versus temperature.

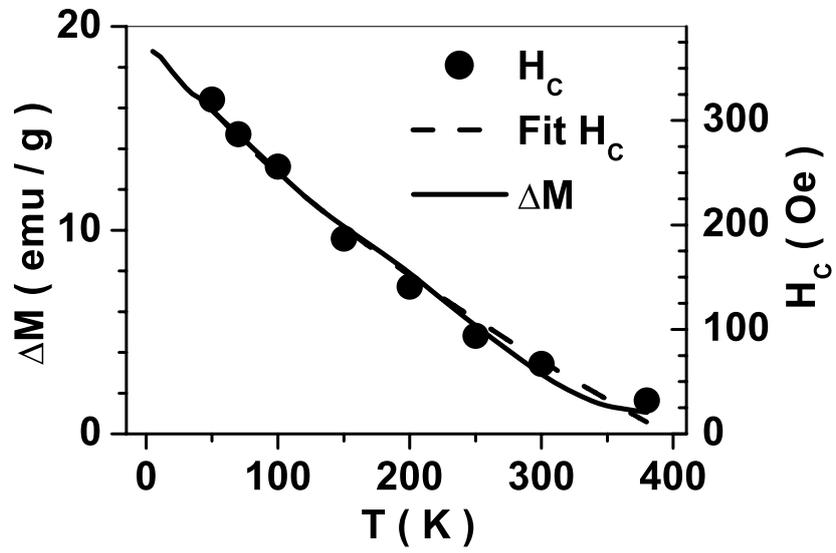

Figure 5: Temperature variation of $H_C(T)$ and $\Delta M(T) = M_{FC} - M_{ZFC}$. The dashed curve is for the fitting result by Eq. (1). The solid curve is for $\Delta M$.